\documentclass[twocolumn,twoside,slac_two]{revtex4}
\usepackage{graphicx}
\usepackage{fancyhdr}
\begin{document}

\title{Broadband multi-wavelength campaign on PKS 2005-489}

%

\author{S. Kaufmann, M. Hauser, S. Wagner}
\affiliation{Landessternwarte, ZAH, Universit\"at Heidelberg, K\"onigstuhl, D69117 Heidelberg, Germany}
\author{M. Raue, O. Tibolla, F. Volpe}
\affiliation{Max-Planck-Institut f\"ur Kernphysik, Heidelberg, P.O. Box 103980, D69029 Heidelberg, Germany}
\author{K. Kosack}
\affiliation{IRFU/DSM/CEA, CE Saclay, F-91191 Gif-sur-Yvette, Cedex, France}
\author{on behalf of the H.E.S.S. collaboration}
\affiliation{..\\}
\author{P. Fortin}
\affiliation{ LLR, Ecole Polytechnique, CNRS/IN2P3, Palaiseau, France}
\author{W. McConville}
\affiliation{NASA Goddard Space Flight Center, Greenbelt, MD 20771\\
University of Maryland, College Park, MD 20742}
\author{D.~J. Thompson}
\affiliation{NASA Goddard Space Flight Center, Greenbelt, MD 20771 }
\author{on behalf of the Fermi/LAT collaboration}
\affiliation{}

\begin{abstract}
The spectral energy distribution (SED) of high-frequency peaked BL Lac objects (HBL) is characterized by two peaks: one in the UV-X-ray and one in the GeV-TeV regime. An interesting object for analyzing these broadband characteristics is PKS 2005-489, which in 2004 showed the softest TeV spectrum ever measured. In 2009, a multi-wavelength campaign has been conducted with, for the first time, simultaneous observations by H.E.S.S. (TeV), Fermi/LAT (GeV), RXTE (keV), Swift (keV, UV, optical) and ATOM (optical) to cover the two peaks of the SED. During this campaign PKS 2005-489 underwent a high state in all wavebands which gives the opportunity to study in detail the emission processes of a high state of this interesting HBL.

\end{abstract}

\maketitle

\thispagestyle{fancy}


\section{Multi-wavelength campaign 2009}
\begin{figure}[h]
\centering
\includegraphics[width=\columnwidth]{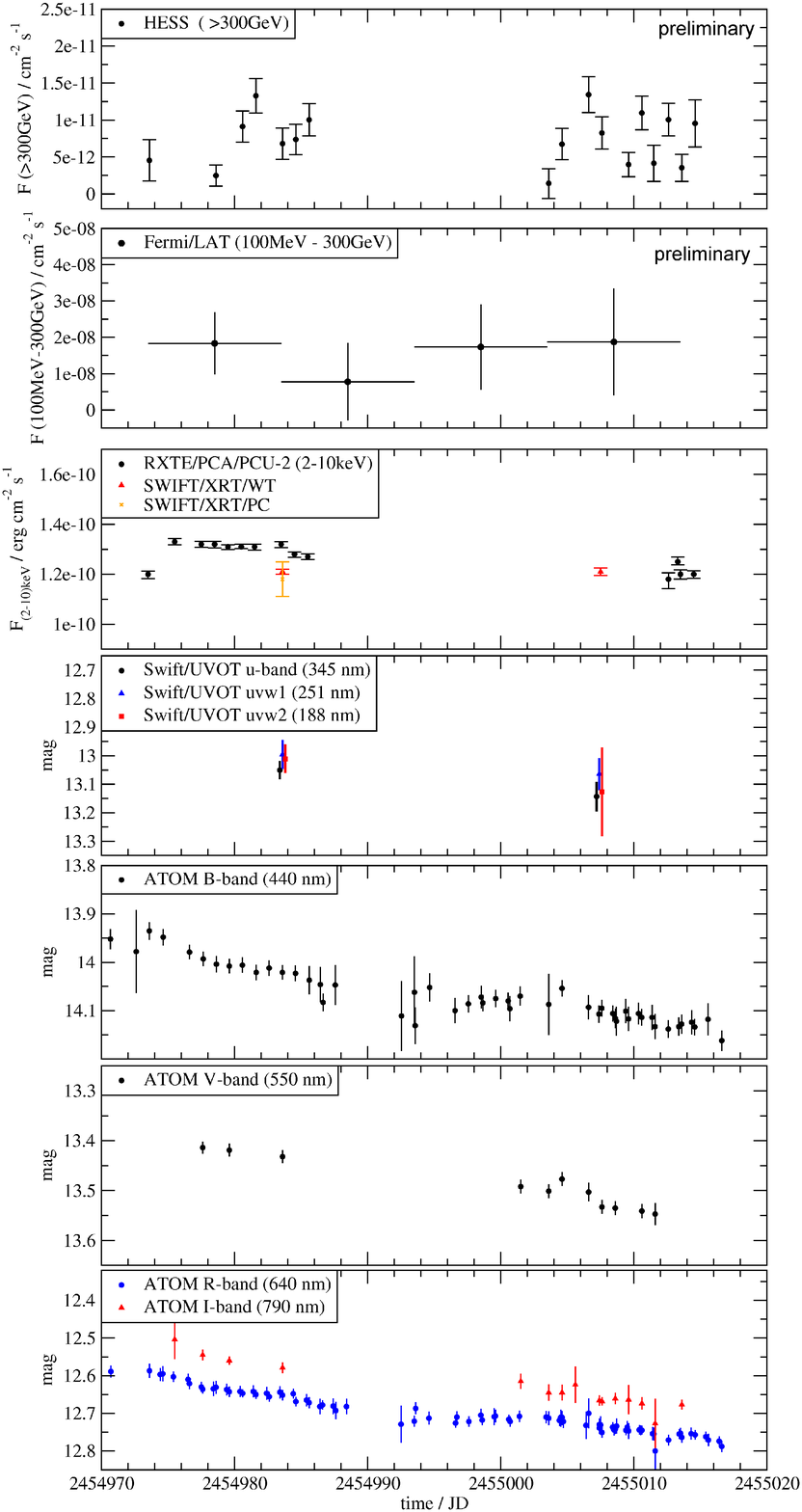}
\caption{Preliminary lightcurves of PKS 2005-489 during this multi-wavelength campaign from 22 May to 2 July 2009 with observations by the Cherenkov telescope array H.E.S.S., the Gamma-ray space telescope Fermi, the X-ray satellites RXTE and Swift and the 75cm optical telescope ATOM. For a better display, the u and uvw2 band data are shifted by 0.2 days. For the H.E.S.S., RXTE and Swift lightcurve a nightly binning was used and the Fermi/LAT lightcurve is shown in a 10-days binning.} 
\label{MWL_LC}
\end{figure}

PKS 2005-489 is one of the brightest high-frequency peaked BL Lac objects (HBL) in the southern Hemisphere. It is located at $\alpha_{\rm{J2000}} = 20^{\rm{h}} 9^{\rm{m}} 25.4^{\rm{s}}$, $\delta_{\rm{J2000}} = -48^\circ 49' 54''$ \cite{Johnston1995} and has a redshift of $z=0.071$ \cite{Falomo1987}. 
In 2004, the first very high energy (VHE) $\gamma$-rays have been detected from PKS 2005-489 by the Cherenkov telescope array H.E.S.S. and the resulting $\gamma$-ray spectrum was the softest TeV spectrum ever measured \cite{Aharonian2005}.
In multi-year studies, PKS 2005-489 show large flux and spectral variations in the X-ray regime, while only weak variation in the VHE band have been detected (e.g. \cite{Aharonian2009}). Therefore simultaneous observations over a broad wavelength range are necessary to investigate the nature of the underlying emission processes.\\
\\
A complex broadband multi-wavelength campaign on PKS 2005-489 was organized by the H.E.S.S. collaboration  and was conducted from 22 May to 2 July 2009 with observations by the Cherenkov telescope array H.E.S.S., the X-ray satellites RXTE and Swift and the optical 75cm telescope ATOM. The LAT instrument onboard the Fermi Gamma-ray Space Telescope is scanning the whole sky within approximately 3 hours and therefore PKS 2005-489 was regularly monitored during this campaign. Therefore simultaneous information about the brightness and the spectrum of the GeV $\gamma$-ray emission could be obtained. 
Hence, for the first time, simultaneous observations have been taken on PKS 2005-489 with H.E.S.S. (TeV), Fermi/LAT (GeV), RXTE (X-ray), Swift (X-ray, UV, optical) and ATOM (optical) that can be used for variability and spectral studies.
The lightcurves resulting from this campaign are shown in Fig.~\ref{MWL_LC}.

\subsection{TeV $\gamma$-ray observations by H.E.S.S.}
The H.E.S.S. experiment consists of four imaging atmospheric Cherenkov telescopes, located in the Khomas Highland, Namibia \cite{Aharonian2006}. The observations on PKS 2005-489 have been taken from 22 May to 2 July 2009 with a break over full moon because the Cherenkov telescopes cannot observe during moontime. The data have been calibrated as described in \cite{Aharonian2004} and analysed using the {\it standard cuts} and following the method described in \cite{Aharonian2006}. After the standard quality selection 13 h of live time remain. A significance of $16\sigma$ (following the method of \cite{LiMa}) result for PKS 2005-489 within the whole observing period. The {\it Reflected-Region} method \cite{Berge2007} was used to determine an appropriate background for the spectrum. An independent analysis has been performed on the same data, but with different calibration, resulting in compatible results.
No significant variation of the very high energy emission is measured on nightly binning during this campaign and a mean flux of $F(>\rm{300GeV}) = 7.9 \times 10^{-12}\;\rm{cm^{-2}\; s^{-1}}$ can be determined. Considering the two months of observations each, also no variability could be detected on month time scales. The measured flux level is $\sim 3$ times brighter than during the detection of this source in 2004 by H.E.S.S. \cite{Aharonian2005}. 

\subsection{GeV $\gamma$-ray observations by Fermi/LAT}
The Large Area Telescope (LAT) onboard the Fermi Gamma-ray Space Telescope is a pair conversion $\gamma$-ray detector sensitive to photons in the energy range from below 20 MeV to more than 300 GeV (\cite{Atwood2009}). 
The data were analyzed using the standard likelihood tools distributed with the Science Tools v9r15p3 package available from Fermi Science Support Center. Only events having the highest probability of being photons, those in the "Diffuse" class, were used. To limit contamination by Earth albedo gamma rays which are produced by cosmic rays interacting with the upper atmosphere, only events with zenith angles $< 105^\circ$ were selected. Time intervals during which the rocking angle was larger than $47^\circ$ were excluded since the Earth's limb was in the field of view. Events with energy between 200 MeV and 300 GeV, and within a $10^\circ$ region of interest (ROI) centered on PKS 2005-489 were analyzed with an unbinned maximum likelihood method (\cite{Cash1979}; \cite{Mattox1996}). The background emission was modeled using a standard galactic diffuse emission  
model and an isotropic component. Point sources within the ROI were also included in the model. The fluxes were determined using the post-launch instrument response functions P6\_V3\_DIFFUSE.
A point source is detected with high statistical significance at more than 20 standard deviations. The best-fit position (20h09m25.0s, $-48^\circ$ 49'44.4'') has a $95\%$ containment radius of 1.5' and is statistically consistent with the coordinates of PKS 2005-489.
The observations of Fermi/LAT on PKS 2005-489 during this campaign show no significant variation and can be described by a mean flux of $F(200\rm{MeV}-300\rm{GeV}) = (1.6 \pm 0.5)\times 10^{-8} \;\rm{cm^{-2}\; s^{-1}}$. Due to the rather faint GeV emission, the binning of the lightcurve for the time of this campaign was choosen to be 10 days. 

\subsection{X-ray observations by RXTE, Swift/XRT}
X-ray observations with the PCA detector onboard RXTE \cite{Bradt1993} have been obtained in the energy range $2-60\;\rm{keV}$ strictly simultaneous to the H.E.S.S. observations from 22 May to 3 June with exposures of $2-4$ ks.
Due to the high state of PKS 2005-489 during this campaign, additional ToO observations have been taken with the X-ray satellites Swift and RXTE. The XRT detector \cite{Burrows2005} onboard Swift observed in photon-counting (PC) and windowed-timing (WT) mode in the energy range $0.2-10\;\rm{keV}$ on 1 June and 24 June with $\sim 3$ks each. The RXTE ToO observations have been performed from 30 June to 3 July.
Only RXTE/PCA data of PCU2 and the top layer have been taken into account to obtain the best signal-to-noise ratio. The data were filtered to account for the influence of the South Atlantic Anomaly, tracking offsets, and electron contamination using the standard criteria recommended by the RXTE GOF. For the count rate of $<40\;\rm{cts/s}$, the faint background model provided by the RXTE GOF was used to generate the background spectrum with {\tt pcabackest} and the response matrices were created with {\tt pcarsp}.
Spectra of the Swift data in pc-mode have been extracted with {\tt xselect} from an annulus region with an outer radius of $0.8'$ at the position of PKS 2005-489, which contains $90\%$ of the PSF at 1.5 keV and an inner radius of $\sim 0.1'$ to avoid the pileup. An appropriate background was extracted from a circular region nearby the source with radius of $3'$. For the wt-mode, appropriate boxes ($\sim 1.6'\times 0.3'$) around the region with source photons and a background region with similar size was used to extract the spectra. 
The auxiliary response files were created with {\tt xrtmkarf} and the response matrices were taken from the Swift package of the calibration database {\tt caldb}.
To determine the flux in the energy range 2-10 keV, a power law model taking into account the galactic absorption of $3.94\times10^{20}\;\rm{cm^{-2}}$ (LAB Survey, \cite{Kalberla2005}) was fitted to the RXTE and Swift spectra. No significant change in spectral shape was found over the time of the campaign.
Slight variations could be seen in the measured flux with an increase at the beginning of the campaign with a following constant flux level over days and a slight decrease at the end of the campaign. The high X-ray flux of PKS 2005-489 is comparable to the historical maximum of 1998 (\cite{Perlman1999}, \cite{Tagliaferri2001}). 

\subsection{UV observations by Swift/UVOT}
The UVOT instrument \cite{Roming2005} onboard Swift measured strictly simultaneous to the X-ray telescope the UV and optical emission in the bands u (345 nm), uvw1 (251 nm) and uvw2 (188 nm) with an exposure of $\sim 1$ ks each. The instrumental magnitudes and the corresponding flux are calculated with {\tt uvotmaghist} taking into account all photons from a circular region with radius $5''$ (standard aperture for all filters). An appropriate background was determined from a circular region with radius $40''$ near the source region without contamination of sources.

\subsection{Optical observations by ATOM}
The 75-cm telescope ATOM \cite{Hauser2004}, located at the H.E.S.S. site in Namibia, monitored the flux in the 4 different filters B (440 nm), V (550 nm), R (640 nm) and I (790 nm). The obtained data have been analysed with the program {\tt midas} using differential photometry with closeby reference stars to determine the absolute magnitudes. At the beginning of the campaign PKS 2005-489 was in a high state and during the campaign the optical flux decreased by $\sim20\%$. Regular monitoring by ATOM and the optical telescope ROTSE show that PKS 2005-489 has a long-term variation of the time scale of years. The monitoring data by ROTSE also shows that the flux in 2009 is the highest flux since 2005. Since the maximum in 2005, the source flux decreased over $\sim 1.5$ years with a slow increase in the following years, until 2009, back to the flux level of 2005. In addition this source shows a smaller short term variation of the optical emission on time scales of months.

\begin{figure}[t]
\centering
\includegraphics[width=\columnwidth]{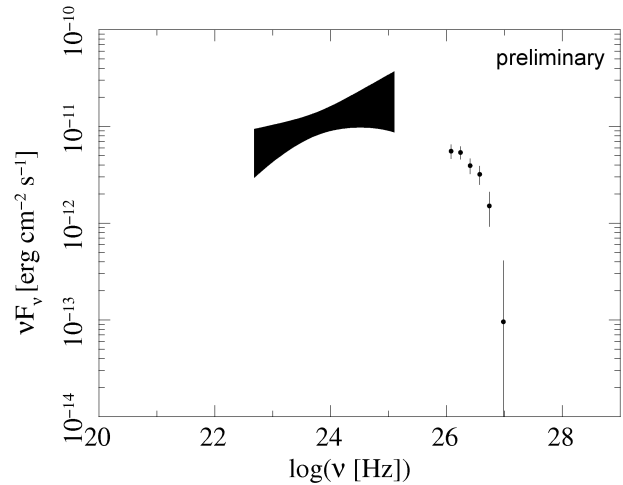}
\caption{Broadband $\gamma$-ray spectrum of PKS2005-489 during this multi-wavelength campaign with simultaneous observations by H.E.S.S. and Fermi/LAT using the time range shown in Fig. 1.} 
\label{SED}
\end{figure}

\section{Broadband spectra}
A major result of this campaign is the good coverage of the high energy peak in the spectral energy distribution by the HE and VHE $\gamma$-ray spectra from the simultaneous Fermi/LAT and H.E.S.S. observations. According to the common leptonic synchrotron self-Compton (SSC) scenario (e.g. \cite{Marscher1985}), this peak can be explained by inverse Compton emission from a population of relativistic electrons which upscatter their own synchrotron photons into the GeV - TeV regime.

Since no significant variation of the high energy radiation was detected by H.E.S.S. and Fermi/LAT, the spectrum for the full time range (as shown in Fig. 1) was created and the resulting $\nu F_\nu$ graph is shown in Fig. 2. Both spectra have been independently fitted by a power law model which is a good description for both energy ranges.
During this campaign PKS 2005-489 shows a harder TeV spectrum than during its TeV detection in 2004 \cite{Aharonian2005} where the softest spectrum of a TeV blazar with spectral index of $\Gamma = 4.0 \pm 0.4$ was measured.
The peak of the inverse Compton emission is located around $10^{25}\; \rm{Hz}$.

\bigskip 
\begin{acknowledgments}
The support of the Namibian authorities and of the University of Namibia in facilitating the construction and operation of HESS is gratefully acknowledged, as is the support by the German Ministry  for Education and Research (BMBF), the Max Planck Society, the French Ministry for Research, the CNRS-IN2P3 and the Astroparticle Interdisciplinary Programme of the CNRS, the U.K. Science and Technology Facilities Council (STFC), the IPNP of the Charles University, the Polish Ministry of Science and Higher Education, the South African Department of Science and Technology and National Research Foundation, and by the University of Namibia. We appreciate the excellent work of the technical support staff in Berlin, Durham, Hamburg, Heidelberg, Palaiseau, Paris, Saclay, and in Namibia in the construction and operation of the equipment. \\
\\
The \textit{Fermi} LAT Collaboration acknowledges generous ongoing support
from a number of agencies and institutes that have supported both the
development and the operation of the LAT as well as scientific data analysis.
These include the National Aeronautics and Space Administration and the
Department of Energy in the United States, the Commissariat \`a l'Energie Atomique
and the Centre National de la Recherche Scientifique / Institut National de Physique
Nucl\'eaire et de Physique des Particules in France, the Agenzia Spaziale Italiana
and the Istituto Nazionale di Fisica Nucleare in Italy, the Ministry of Education,
Culture, Sports, Science and Technology (MEXT), High Energy Accelerator Research
Organization (KEK) and Japan Aerospace Exploration Agency (JAXA) in Japan, and
the K.~A.~Wallenberg Foundation, the Swedish Research Council and the
Swedish National Space Board in Sweden.

Additional support for science analysis during the operations phase is gratefully
acknowledged from the Istituto Nazionale di Astrofisica in Italy and the Centre
National d'\'Etudes Spatiales in France.\\
\\
The authors acknowledge the support by the RXTE and Swift teams for providing ToO observations and the use of the public HEASARC software packages. 

\end{acknowledgments}

\bigskip 

\end{document}